\documentclass[twocolumn,pra]{revtex4}
\usepackage{graphicx}
\usepackage{epsfig}
\usepackage{latexsym}
\usepackage{amssymb}
\usepackage{bm}
\usepackage{float}
\usepackage{subfigure}
\usepackage{amsmath}
\usepackage{amsfonts}
\begin{document}

\title{Killing the Straw Man: Does BICEP Prove Inflation at the GUT Scale?}
\author{James B. Dent}
\affiliation{Department of Physics, University of Louisiana at Lafayette, Lafayette, LA 70504, USA,}
\author{Lawrence M. Krauss}
\affiliation{Department of Physics and School of Earth and Space Exploration, Arizona State University, Tempe, AZ 85287, USA, and Mount Stromlo Observatory, Research School of Astronomy and Astrophysics, Australian National University, Weston, ACT, Australia, 2611}
\author{Harsh Mathur}
\affiliation{Department of Physics, Case Western Reserve University, Cleveland, Ohio 44106-7079}

\begin{abstract}

The surprisingly large value of $r$, the ratio of power in tensor to scalar density perturbations in the CMB reported by the BICEP2 Collaboration, if confirmed, provides strong evidence for Inflation at the GUT scale.    While the Inflationary signal remains the best motivated source, a large value of $r$ alone would still allow for the possibility that a comparable gravitational wave background might result from a self ordering scalar field (SOSF) transition that takes place later at somewhat lower energy.  We find that even without detailed considerations of the predicted BICEP signature of such a transition, simple existing limits on the
isocurvature contribution to CMB anisotropies would definitively rule out a contribution of more than $5\%$ to $r \approx 0.2$,.  We also present a general relation for the allowed fractional SOSF contribution to $r$ as a function of the ultimate measured value of $r$.  These results point strongly not only to an inflationary origin of the BICEP2 signal, if confirmed, but also to the fact that if the GUT scale is of order $10^{16} GeV$ then either the GUT transition happens before Inflation or the Inflationary transition and the GUT transition must be one and the same.  
\end{abstract}

\maketitle

The recent claimed observation of primordial gravitational waves \cite{Ade:2014xna}, if confirmed, would provide a dramatic new empirical window on the early universe.  In particular, it would provide the 
opportunity, in principle, to definitively test the inflationary paradigm\cite{Guth:1980zm,Linde:1981mu}, and to explore the specific physics of inflationary models.  However, while there is little doubt
that inflation at the Grand Unified Scale is the best motivated source of such primordial waves (e.g. \cite{Rubakov:1982df,Fabbri:1983us,Abbott:1984fp,Krauss:1992ke}, it is important to demonstrate that other possible early universe sources could not account for the current
BICEP2 data, if validated, before definitely claiming Inflation has been proved.  

A surprisingly large value of $r$, the ratio of power in tensor modes to scalar density perturbations provides a challenge for other possible primordial
sources.  Here we utilize a simple and robust constraint on such sources:  They would have to generate gravitational waves efficiently without altering the observed adiabatic density fluctuations that are so consistent with
inflationary predictions.  Our analysis also allows a determination of the possible fractional contribution to $r$ from such scenarios as a function of the ultimate measured value of $r$.   Moreover, we point out another important implication of the BICEP result, if it is validated.  It implies that the Inflation scale and the GUT scale must in general be coincident, or the GUT transition must occur before Inflation.

We have previously explored a relatively generic possible competing source of a scale invariant spectrum of tensor modes \cite{Krauss:1991qu,JonesSmith:2007ne,Krauss:2010df}, a simple self ordering scalar field (SOSF) in the early universe, 
 and one might hope that the BICEP2 observation would
rule out this possibility, as well as other ones involving causal processes inside the horizon (see \cite{Stebbins:1987va,Pen:1993nx,Turok:1997gj,Seljak:1997ii,Pen:1997ae} for key results relevant to the impact of a such processes on scalar, vector and tensors modes in the CMB,) thus allowing a cleaner interpretation of the the existing data in terms of inflation.  Indeed, we demonstrate that existing bounds on any possible isocurvature component in
the scalar power spectrum rule out the possibility of any significant SOSF contribution to the BICEP2 observation.   This implies the BICEP2 result most likely does reflect gravitational waves from inflation, with all of the exciting concomitant implications (i.e. quantization
of gravity \cite{Krauss:2013pha}).  

In the following we assume inflation occurs, and provides the measured adiabatic scalar density fluctuations inferred from CMB measurements (because that is strongly suggested by the data), but that a SOSF phase transition occurs after inflation, producing a gravitational wave signature that might overwhelm the inflationary signal. 

Let $S_i$ and $T_i$ denote the scalar and tensor power generated by inflation and $S_\varphi$ and $T_\varphi$
the same quantities for the self-ordered scalar field. Out of these four quantities one can form several ratios of
interest: (i) $r_{{\rm eff}} = (T_i + T_\varphi)/(S_i + S_\varphi)$ is the tensor to scalar ratio incorporating both sources that has just been
observed to have a central value of 0.2. (ii) The self-ordering scalar field produces isocurvature scalar fluctuations whereas inflation
produces adiabatic ones. Measurements of the temperature anisotropies constrain the isocurvature fraction
$x = S_\varphi/(S_i + S_\varphi)$ to lie in the range $0 < x < 0.09$ \cite{Ade:2013uln}.  Note that we are taking the {\it{most conservative upper bound}} on the isocurvature modes. A detailed calculation including constraints on the scale dependence of isocurvature modes in the SOSF framework would improve this bound.  However, as we shall show, even this conservative bound is sufficient to rule out a significant SOSF contribution to the BICEP2 signal. (iii) $r_\varphi = T_\varphi/S_\varphi$, the tensor-to-scalar ratio for the SOSF case, can be
calculated within the self-ordering scalar field model using the scalar power spectrum described in \cite{Figueroa:2010zx} along with the tensor power given in \cite{JonesSmith:2007ne,Fenu:2009qf,Krauss:2010df}, and is found to be 0.118 \footnote{The first calculation of the tensor power was done in \cite{JonesSmith:2007ne}.  The normalization of the calculation was found to be in error as noted in \cite{Fenu:2009qf}, and subsequently corrected in \cite{Krauss:2010df}.   Its value is $(\kappa/(4\pi^5))(\eta/(N^{1/4}M_{Pl})^4$, where $\kappa = 11600$, while the scalar power is given by $(\eta/M_{Pl})^4(80/N)$, which gives the ratio of about 0.118.} 
(iv) $f = T_\varphi/T_i$, the ratio of the tensor contributions from the SOSF mechanism to that produced by inflation, is given by
$(140/N) (V_\varphi/V_i)$ \cite{JonesSmith:2007ne,Fenu:2009qf,Krauss:2010df} where $N$ denotes the number of components of the self-ordering scalar field
(presumed to be large and definitely greater than three), $V_\varphi$ is the symmetry breaking scale for the
self-ordering field and $V_i$ is the scale of inflation. We need $V_\varphi < V_i$ to ensure that symmetry breaking
occurs after inflation (otherwise evidence of it would be obliterated by inflation). This inequality constrains the
ratio $f$. (v) The tensor to scalar ratio for inflation $r_i = T_i/S_i$ is the quantity of interest for inflationary models. 
In the absence of the self-ordering scalar fields, $r_i$ is equal to the measured quantity $r_{{\rm eff}}$, but   $r_i$ could have a considerably lower value if
self-ordering scalar fields dominated the observed signal. 

\begin{figure}[h]
\begin{center}
\includegraphics[width=0.4\textwidth]{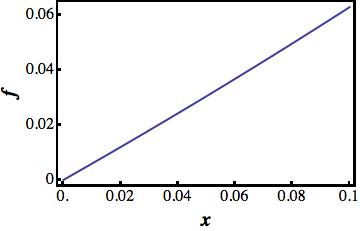}
\caption{Plot of $f$, the ratio of tensor contributions from SOSF to those of inflation as a function of the isocurvature fraction, $x$.
}
\label{fig:fplot}
\end{center}
\end{figure}

Since only three of these ratios are independent,
but there are now constraints on four of them, in principle, the data is capable of ruling out the
existence of self-ordering scalar fields as a source. To explicitly determine the constraints we express $f$ in terms
of $r_{{\rm eff}}, x$ and $r_\varphi$
\begin{equation}
f = \frac{x r_\varphi}{r_{{\rm eff}} - x r_\varphi}.
\label{eq:f}
\end{equation}
Fig.\ref{fig:fplot} shows a plot of $f$ as a function of $x$ reveals that $f$ grows monotonically with $x$, and for the conservative upper limit on $x$ of $0.09$, the maximal fractional contribution of SOSF to BICEP2 is less than about $5\%$.  (Note that if the Planck foreground dust polarization maps are given the most weight, the central value of $r$ inferred by BICEP reduces to $r=0.16$.  In this case, the maximum contribution of SOSF would rise to about $7\%$.)

\begin{figure}[h]
\begin{center}
\includegraphics[width=0.4\textwidth]{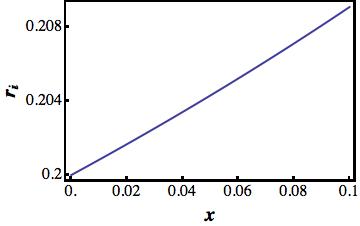}
\caption{The inflationary tensor-to-scalar ratio, $r_i$ as a function of the isocurvature fraction, $x$.}
\label{fig:riplot}
\end{center}
\end{figure}

It is also worth noting that because $r_{\phi} < 1$, the contribution of SOSF's is greater to scalar density modes than it is to tensor modes.  This means that if one allows for a non-zero SOSF contribution to the BICEP2 result, in order for $r_{eff}=0.2$, the contribution of Inflationary modes $r_i$ would actually need be greater than 0.2, increasing in proportion to the size of the SOSF contribution to BICEP, as can be seem in Fig. \ref{fig:riplot}.

We also note that the current analysis has not included the possible contribution from vector modes due to SOSF.  However since such modes are known to contribute roughly equally to scalar and tensor modes in the CMB it should not significantly affect ratios.  Although it would need to be calculated and included in a more complete future quantitative analysis, the strength of the constraint on SOSF contributions to the BICEP2 is sufficiently significant so that no qualitative change and very little quantitative change might be expected. 

Finally we note that the constraint on SOSF's actually puts a severe constraint on non-inflationary phase transitions that might happen near, but below the Inflation scale.  In particular a GUT transition involving symmetry breaking of a non-abelian gauge symmetry would be expected to produce tensor and scalar perturbations of magnitude comparable to that quoted here for our toy SOSF model.  This means that in order for the GUT scale to be comparable to the Inflation scale, the GUT transition must either occur before Inflation, with a larger scale than that associated with the Inflation scale, or the GUT transition and the Inflationary transition need to be one and the same.    Thus, the fact that our analysis provides additional evidence that the BICEP2 result arises from inflation and not from SOSFs, also allows significant new constraints on the scale and nature of Grand Unification.

\emph{Note added:} In this work we focused on the overall magnitude of
the claimed signal, independent of its spectral features, and argue that the magnitude alone rules out SOSF model through conflict with known constraints on isocurvature contributions to the CMB.
After completion of an initial version of this work Durrer {\it et al} \cite{Durrer:2014raa}
have performed a detailed comparison of the
the observed $C_{\ell}^{BB}$ power spectrum of B-modes to that calculated within the
SOSF model as first done by Garcia-Bellido et al. \cite{GarciaBellido:2010if}.
They find that if only the data at low-$\ell$ are considered, a SOSF model alone provides a poor direct fit to the data compared to the inflationary prediction because in the former case super horizon effects are uncorrelated, causing the power spectrum to fall off for small $\ell$ whereas the signal apparently does not. A similar conclusion applies to the signal from models involving topological defects, for the same reason \cite{Lizarraga:2014eaa}. Note that this conclusion depends strongly not weighting the higher $\ell$ data points from BICEP2, which do not fit the Inflationary prediction and which the collaboration themselves suggest should not be as strongly weighted.
A consideration of the low-$\ell$ data then allow Durrer {\it et al}  to
put direct constraints on the SOSF model contribution from the BICEP2 data alone.  These are consistent with the upper bound we derive here based solely a consideration of the isocurvature contribution from SOSFs. Together these results strongly reinforce the likely conclusion
that the observed B-mode signal, if confirmed 
is due to inflation.

We acknowledge discussions with Kate Jones-Smith and Paul J. Steinhardt at an early stage of this work, and we thank referees for pointing out a numerical error in our initial analysis which affected our conclusions.

\end{document}